\documentstyle[aps,amssymb]{revtex}

\newcommand{\nc}{\newcommand}
\nc{\vare}{\varepsilon}
\nc{\erm}{{\rm e}} \nc{\dis}{\displaystyle}
\nc{\vecna}{\mbox{\boldmath $\nabla$}}
\nc{\pa}{\partial} \nc{\ug}{\; = \;} \nc{\vs}{\vspace*}
\nc{\rd}{\rm d}
\nc{\imp}{\mbox{\boldmath $p$}} \nc{\vbf}{\mbox{\boldmath $v$}}
\nc{\abf}{\mbox{\boldmath $a$}} \nc{\Fbf}{\mbox{\boldmath $F$}}

\def\be{\begin{equation}}
\def\ee{\end{equation}}
\def\bea{\begin{eqnarray}}
\def\eea{\end{eqnarray}}
     
\nc{\bx}{{\bf x}}

\begin{document}

\title{LABORATORY BOUNDS ON LORENTZ SYMMETRY VIOLATION IN LOW ENERGY
NEUTRINO PHYSICS$^{\star}$}

\footnotetext{$\!\!\!\!^{\star}\,$Work partially supported by I.N.F.N.
and M.I.U.R.}

\maketitle
\begin{center}
{\bf E. Di Grezia}, \ {\bf S. Esposito}

\noindent {\em Dipartimento di Scienze Fisiche, Universit\`a di
Napoli ``Federico II''}\\
and\\
{\em Istituto Nazionale di Fisica Nucleare, Sezione di Napoli \\
Complesso Universitario di Monte S. Angelo,  Via Cinthia, I-80126
Naples, Italy
\\
{\rm e-mail:} digrezia@na.infn.it; sesposito@na.infn.it}

\

{\bf G. Salesi}

{{\em Universit\`a Statale di Bergamo, Facolt\`a di
Ingegneria,\\
viale Marconi 5, 24044 Dalmine (BG), Italy}\\
and\\
{\em Istituto Nazionale di Fisica Nucleare, Sezione di Milano\\
via Celoria 16, I-20133 Milan, Italy\\
{\rm e-mail}: salesi@unibg.it}}
\end{center}


\begin{abstract}
\noindent Quantitative bounds on Lorentz symmetry violation in the
neutrino sector have been obtained by analyzing existing
laboratory data on neutron $\beta$ decay and pion leptonic decays.
In particular some parameters appearing in the energy-momentum
dispersion relations for $\nu_e$ and $\nu_\mu$ have been
constrained in two typical cases arising in many models accounting
for Lorentz violation.

\end{abstract}

\section{Lorentz-violating dispersion relations}

\noindent In recent times ultra-high energy Lorentz symmetry violations
have been investigated, both theoretically and
experimentally,\footnote{Hereafter for simplicity we use the term
``violation'' or ``breakdown'' of the Lorentz symmetry, but, in
some theories, although Special Relativity does not hold anymore,
an underlying extended Lorentz invariance exists (this happens,
for example, in ``Doubly'' Special Relativity\cite{ChenYang,GAC},
where ``deformed'' 4-rotation generators are considered).} by means of
quite different approaches, sometimes extending, sometimes abandoning
the formal and conceptual framework of Einstein's Special Relativity.
The most important consequence of a Lorentz violation (LV) is the
modification of the ordinary momentum-energy dipersion law
 \ $E^2=\imp^2+m^2$ \ at energy scales usually assumed of the
order of the Planck mass, by means of additional terms which vanish
in the low momentum limit. Lorentz-breaking observable effects appear
in Grand-Unification Theories\cite{GUTs}, in (Super)String/Brane
theories\cite{String}, in Quantum Gravity\cite{QG}, in 
foam-like quantum spacetimes\cite{Foam}; in spacetimes endowed with a
nontrivial topology or with a discrete structure at the Planck
length\cite{Spacetimes,Alfaro}, or with a (canonical or noncanonical)
noncommutative geometry\cite{ChenYang,GAC,NCG}; in the so-called
``extensions'' of the Standard Model incorporating breaking of Lorentz
and CPT symmetries\cite{SME}; in theories with a variable speed of light
or variable physical constants.
In particular, the M-Theory\cite{String} and the Loop Quantum Gravity
(as in the version with semiclassical spin-network structure of
spacetime)\cite{Spacetimes,Alfaro,LQG} lead to postulate an
essentially discrete, quantized spacetime, where a fundamental
mass-energy scale naturally arises, in addition to $\hbar$ and
$c$. An intrinsic length is directly correlated to the existence
of a ``cut-off'' in the transferred momentum necessary to avoid
the occurrence of ``UV catastrophes'' in Quantum Field Theories.
Other divergences, as those emerging in black hole entropy, could
possibly be averted in the presence of quantum LV's.

A natural extension of the standard dispersion law can be put in
most cases under the general form ($p\equiv|\imp|$) \be E^2 =
p^2+m^2+p^2f(p/M)\,, \label{s2} \ee where $M$ indicates a (large)
mass scale characterizing LV. By using a series expansion for $f$,
under the assumption being $M$ a very large quantity,
we can consider only the lower order nonzero term in the
expansion: \be E^2 = p^2+m^2+\alpha
p^2\left(\frac{p}{M}\right)^n\,. \ee The most recurring exponent
in the literature on LV is $n=1$: \be E^2 =
p^2+m^2+\alpha\frac{p^3}{M}\,. \label{eq:cubic} \ee Indeed in
``noncritical''-Liouville String Theory\cite{Ellis}, we have \
$E^2 = p^2+m^2+\xi g_sp^3/M_s$, \ where $g_s$ is the string
coupling and $M_s$ is a string (non-Planck) mass scale; similar
expressions are obtained in the Standard Model extensions of
Colladay and Kosteleck\'y, as well as in theories with a spacetime
``medium'' or quantum foam, or even in Quantum Gravity. We find
dispersion laws analogous to the $n=1$ case in Deformed or Doubly
Special Relativity,\cite{GAC,NCG,Deformed} working in k-deformed
Lie-algebra noncommutative (k-Minkowski) spacetimes, in which both
the Planck scale and the speed of light act as characteristic
scales of a 6-parameter group of spacetime 4-rotations with
deformed but preserved Lorentz symmetries. For example, in
\cite{GAC} where the Lie algebra is given by \ $[x_i,x_0]=il x_i$,
\ $[x_i,x_k]=0$ \ ($l$ being a very small length which may
eventually be taken of the order of $M_{\rm Planck}^{-1}$) we have
\be \frac{\erm^{lE}+\erm^{-lE}-2}{l^2}- p^2\erm^{-lE} = m^2 \ee
which for $p\gg m$ and considering terms up to O($l^2$) reduces
to the above seen cubic form \be E^2=p^2+m^2\mp l
p^3\,. \ee The $n=0$ case in (\ref{s2}) has been studied by
Coleman and Glashow\cite{Coleman}, while $n=2$ is found for some
``Quantum Deformed Poincar\'e Groups''\cite{NCG}, in spacetime
foam scenarios at low energies\cite{Foam,Ellis} and in effective
field theories including Lorentz violating dimension-5
operators\cite{Jacobson,Myers}. Different dispersion laws arise
from canonical Loop Gravity, Supergravity, or String Theory
critical in B-background\cite{Critical,GAC3}, and from canonical
noncommutative geometries\cite{GAC}. For example in \cite{GAC},
assuming the canonical commutation relation
 \ $[x^\mu, x^\nu]=i\theta^{\mu\nu}$ \, we find
\be E^2=p^2+m^2+\frac{\eta}{(p_\mu\theta^{\mu\nu})^2}\,. \ee Other
important models, which assume the Planck 4-momentum (a new
Lorentz invariant besides the speed of light) as a
frame-independent quantity, entail a linear behavior (i.e.,
$n=-1$): in \cite{Alfaro,NCG,Smolin,Majid,CC} for $p\gg m$ and up
to terms of the order of $\mu^2$, $\mu$ driving the mass scale of
the Lorentz breakdown, we have \be E^2=p^2+m^2 \mp 2\mu p.
\label{eq:linear} \ee In such models it is interesting to study
the (small momenta) nonrelativistic limit wherein the nonstandard
linear term rules.

As pointed out by Bertolami\cite{Bertolami}, evidences of violation
of the Lorentz symmetry seem to emerge from the observation of: a)
ultra-high energy cosmic rays with energies\cite{UHECR} beyond the
Greisen-Zatsepin-Kuzmin\cite{GZK} cut-off (of the order of
$4\times 10^{19}$\,eV); \ b) gamma rays with
energies beyond 20 TeV from distant sources such as Markarian 421
and Markarian 501 blazars\cite{Markarian}; \ c) longitudinal
evolution of air showers produced by ultra-high energy hadronic
particles which seem to suggest that pions are more stable than
expected\cite{Showers}.

The theoretical applications of the modified dispersion relation
do carry many threshold effects associated to asymmetric momenta
in photoproduction, pair creation, photon stability, vacuum
\v{C}erenkov effects, etc.\cite{Cerenkov,Jacobson2}, long baseline
dispersion and vacuum birefringence (signals from gamma ray
bursts, active galactic nuclei, pulsars)\cite{Gleiser}, dynamical
effects of LV background fields (gravitational coupling and
additional wave modes)\cite{Jacobson}, different maximum speeds
for different particles\cite{Coleman,Jacobson}.

An interesting consequence of a nonstandard dispersion law is an
essentially ``non-Newtonian'' dynamics at very high
energies.\footnote{In fact, if \ $E^2 \neq p^2+m^2$, from \
$\vbf=\partial E/\partial\imp$ \ we have not \
$\vbf=\imp/E$. Thus, in four-dimensional notation, we have \ $v^
\mu\neq p^\mu/m$ \ and, time-differentiating both sides,
$a^\mu\neq F^\mu/m$ ($F^\mu=\dot{p}^\mu$ indicating the 4-force):
Newton's Law does not hold anymore.} All that recalls the
non-Newtonian features of spinning particles dynamics at the
Compton scale $\hbar/m$\cite{Salesi}.

There are further deep experimental implications at low
energy\cite{SME,Low,Experiments}, mostly in the Standard Model
extensions based on a Lagrangian containing any possible
phenomenologically relevant Lorentz- and CPT-violating term, for
hadrons\cite{Hadrons}, nucleons\cite{Nucleons}, electrons\cite{Electrons},
photons\cite{Photons}, muons\cite{Muons}.

Consequences on the absorption and the spectrum of gamma rays and
on the synchrotron radiation have been investigated in
noncommutative QED\cite{Castorina} as well.

Allen and Yokoo\cite{Allen} list a series of proposed or performed
experiments, on Earth and in space, to test both Lorentz and CPT
symmetries: atomic experiments\cite{SME,QG,Experiments} (penning
trap experiments with electrons, protons and their respective
antiparticles, clock comparison experiments exploiting Zeeman and
hyperfine transitions, spin polarized torsion pendulum
experiments, etc.); clock-based experiments\cite{Experiments} to
probe effects of variations in both orientation and velocity
(employing H-masers, laser-cooled Cs and Rb clocks, dual nuclear
Zeeman He-Xe masers, superconducting microwave cavity
oscillators); experiments involving neutrino and kaon
oscillations\cite{Experiments}; measurements of cosmological
birefringence by interferometric searches of spacetime metric
fluctuations\cite{GAC2}.

Breakdown of the relativistic invariance and deformed ``mass
shell'' relation have been advanced, above all, for {\em
neutrinos} because of their very small mass and very large
momenta. Consequences of a nonstandard dispersion law on neutrino
oscillations has been analyzed in \cite{ChenYang} and
\cite{Coleman}; the observable effects of the non-Lorentz nature
of the flavour eigenstates have been investigated by Blasone {\em
et al.}\cite{Blasone}, by Klinkhamer\cite{Klink} and by Smolin and
Maguejio\cite{Smolin}. In particular, striking effects on neutrino
physics are expected\cite{SME,Coleman,CC,Barger} as a possible
explanation of the arrival delays of neutrinos emitted from
supernova SN1987A, as well as a possible kinematical stability of
neutrons and pions of very high momentum. As stated in
\cite{GAC3,CC,Jacobson2} SN1987A might constitute an interesting
laboratory for studying LV's, because of the relatively high
energy of the observed neutrinos (up to 100MeV), the relatively
large distances travelled (about $10^4$ light-years), and the
short (of the order of second) duration of the bursts.

By assuming the dispersion relation (\ref{eq:linear}) for the mass
eigenstates of electron neutrinos, Carmona and Cortes\cite{CC}
succeed in explaining the so-called ``tritium beta-decay anomaly",
i.e., the anomalous excess of decay events near the endpoint of
the electron energy spectrum (where nonrelativistic few-eV
neutrinos are produced) which yields a characteristic ``tail'' in
the Kurie plot. If the Special Relativity dispersion law holds,
this unexpected phenomenon involves {\em negative} squared masses
for electron neutrinos. Instead, using eq.\,(\ref{eq:linear}), those
authors show that the excess of electron events is only apparent
and no tachyonic neutrino is needed. In the same paper the Lorentz
violating momentum-energy relation is exploited to account for the
spread of arrival times of neutrinos from SN1987A, and for the
stability of very high energy neutrons and pions in cosmic rays,
since the high momentum decay reactions $n\to
p+e^-+\overline{\nu}_e$, \ $\pi^-\to \mu^-+\overline{\nu}_\mu$ 
\ become forbidden. 

In the next sections we shall deduce upper bounds on the LV
parameters appearing in equations (\ref{eq:cubic}) and
(\ref{eq:linear}), respectively, by using the available precision
experimental data on the neutron lifetime and the decay rates of
the charged $\pi$ meson.

We assume that even if the kinematics of the above decays is
affected by the LV in the dispersion law, nevertheless the
dynamics, at the low energy scales of laboratory experiments, is
essentially the one given by the Standard Model. If other
radiative contributions (as additional Feynman diagrams) should
appear as a consequence of the modified mass shell, we expected 
they are irrelevant at the considered energy scales, due to the
excellent experimental verification of the Standard Model
predictions at these scales.

\section{Neutron $\beta$ decay}

\noindent Let us first consider modifications induced by a Lorentz violating
dispersion relation for neutrinos on the lifetime of the neutron.
The relevant decay channel is: \be n \rightarrow p + e^- +
\bar{\nu}_e\,, \label{1}\ee and the differential decay rate,
assuming that the Standard Model gives the dominant contribution,
in the Born approximation can be written as \cite{Okun}: \be
\rd\Gamma = \frac{eG_{\rm F}^2(C_{\rm V}^2+3C_A^2)}{(2\pi)^4}\,\delta(E_\nu
-Q)p_\nu^2\rd p_\nu p_e^2\rd p_e\rd\Omega_e\,,\label{2}\ee where
$G_{\rm F}$ is the Fermi coupling constant and $C_{\rm V}$, $C_{\rm A}$
are the vector and axial couplings. We have denoted with $E_i, p_i$ the
energy and $3$-momentum of the particle $i$, while $Q= \Delta - E_e$ and
$\Delta$ is the neutron-proton mass difference: $ \Delta  =M_n -
M_p$. As a first case, we consider a neutrino dispersion relation
of the form (\ref{eq:linear}) \be E_\nu^2 = p_\nu^2 + m_\nu^2 +
2\mu p_\nu\,.\label{3}\ee In the ultrarelativistic limit $m_\nu
\ll p_\nu$ (and also assuming that $\mu$ is small compared to
the neutrino momentum) we thus have: \be E_\nu \simeq p_\nu +
\frac{m_\nu^2}{2p_\nu} + \mu \label{4} . \ee For simplicity, in
what follows we take $m_\nu =0$; by inserting (\ref{4}) into
(\ref{2}) and performing some integrations, at first order in
small quantities we obtain: \be \Gamma \simeq \Gamma_{\rm B} + 2 \frac{
\mu_e G_{\rm F}^2(C_{\rm V}^2 + 3C_{\rm A}^2)}{m_e(2\pi)^3}
m_e^5\int_1^{\Delta/m_e}\rd\varepsilon\,\varepsilon
\sqrt{\varepsilon^2 -1}\left(\varepsilon -
\frac{\Delta}{\em m_e}\right), \label{5}\ee where $\varepsilon =
E_e/m_e$, and \be \Gamma_{\rm B} =\frac{G_{\rm F}^2(C_{\rm V}^2 +
3C_{\rm A}^2)}{(2\pi)^3} m_e^5\int_1^{\Delta/m_e}\rd\varepsilon\,
\varepsilon \sqrt{\varepsilon ^2 -1}\left(\varepsilon -
\frac{\Delta}{\em m_e}\right)^2, \label{6}\ee is the standard Born
rate for the neutron decay. An accurate comparison with the
experimental date, in order to get a constraint on the parameter
$\mu_e$, should require the computation of the radiative
corrections (and similar ones) to the Born expressions (see, for
example, \cite{precisium}). However, here and in the following we
are interested only in giving an order of magnitude estimate for
$\mu_e$, as can be deduced from laboratory experiments. To this
end, let us write the total (corrected) neutron decay rate as: \be
\Gamma = \Gamma_0 + 2 \frac{ \mu_e }{m_e}\Gamma_1\,,\label{7}\ee
where $\Gamma_0$ is the (corrected) Standard Model rate while
$2m_e\Gamma_1/\mu_e$ represents the Lorentz-breaking term. The parameter
$\mu_e$ is then obtained from: \be 2 \frac{ \mu_e }{m_e}=
\frac{\Delta\Gamma}{\Gamma_1} =\frac{\Delta\Gamma}{\Gamma_0}\,\,
\frac{\Gamma_0}{\Gamma_1}\,,\label{8}\ee $\Delta\Gamma = \Gamma -
\Gamma_0$. The factor $\Delta\Gamma/\Gamma_0$ is approximately
given by the relative uncertainty in the experimental
determination of the neutron lifetime $\tau$; from the Particle
Data Group analysis we deduce \cite{PDG}: \be
\frac{\Delta\Gamma}{\Gamma_0}\le \frac{\Delta\tau}{\tau}\sim
10^{-3}\,.\label{9}\ee Instead, the factor $\Gamma_0 / \Gamma_1$ can
be roughly estimated as follows. From Eqs. (\ref{5}) and (\ref{6})
we can deduce that \be \frac{\Gamma_0}{\Gamma_1}\sim\left\langle\left|\varepsilon
- \frac{\Delta}{m_e}\right|\right\rangle\,,\label{10}\ee where $\left\langle\varepsilon -
\Delta/m_e\right\rangle$ denotes a sort of  average over the electron
energy spectrum. The kinematics of the neutron beta decay predicts
that a typical average value for the electron energy is of the
order of $1 $ KeV, so that we can estimate: \be
\frac{\Gamma_0}{\Gamma_1} \sim 2\,,\ee that is, such a value is of
the order of unity. From Eq. (\ref{8}) we finally obtain the
following rough limit  on the Lorentz-violating $\mu_e$ parameter
for the electron-neutrino relation: \be \mu_e\le 1 \ {\rm keV}\,.
\label{12}\ee Similar considerations hold when we consider the
dispersion relation in (\ref{eq:cubic}) and replace Eq.(\ref{3})
with the following \be E_\nu^2 = p_\nu^2 + m_\nu^2 + \alpha
\frac{p_\nu^3}{M} . \label{13}\ee  Repeating the computations for
the decay rate along the lines outlined above, Eq.(\ref{5}) is now
replaced by \be \Gamma \simeq \Gamma_{\rm B} + 2 \frac{m_e\alpha_e}{M} \
\frac{G_{\rm F}^2(C_{\rm V}^2 + 3C_{\rm A}^2)}{(2\pi)^3}
m_e^5\int_1^{\Delta/m_e}\rd\varepsilon\,\varepsilon
\sqrt{\varepsilon ^2 -1}\left(\varepsilon -
\frac{\Delta}{\em m_e}\right)^3. \label{14}\ee Now we set: \be  \Gamma
= \Gamma_0 + 2\frac{\alpha_e}{M} m_e \Gamma_1 ,\ee and \be
\frac{\Gamma_0}{\Gamma_1}\sim\left\langle\left|\varepsilon -
\frac{\Delta}{m_e}\right|\right\rangle^{-1}.\label{16}\ee 
We thus obtain the following order of magnitude constraint on the 
$\nu_e$ parameter $\alpha_e$: \be \alpha_e^{-1}M\ge 1 \ {\rm GeV}\,.\ee

\section{Pion leptonic decay}

\noindent Since long time the ratio of the decay rates of the charged $\pi$
meson, \be R_{e\mu}=  \frac{\Gamma (\pi \rightarrow e
\nu_e)}{\Gamma (\pi \rightarrow \mu \nu_\mu)}, \label{21}\ee has
been used to test fundamental properties of elementary particle
physics; this quantity in fact at a first, good approximation, is
independent of the details of $\pi$ meson interactions
\cite{Okun}. Assuming that the dynamics of the considered decays
is substantially described in the framework of the Standard Model,
the ratio in (\ref{21}) can also be useful to obtain constraints
on the Lorentz-breaking parameters in the neutrino dispersion
relations. Note that, on the contrary to what happens for the
neutron decay, the single decay rates (instead of their ratio)
cannot give useful informations on these parameters, since the
experiments measuring directly the above mentioned processes are
used to fix the value of the pion decay constant $f_\pi$
\cite{PDG}. The decay rate of a charged pion into a
lepton-neutrino pair can be written as \cite{Okun}: \be \Gamma
(\pi \rightarrow l \nu)= \frac{G_{\rm F}^2\cos{\theta_c}^2
f_\pi^2m_l^2E_\nu }{2\pi}\int\frac{p_\nu^2\rd p_\nu}{E_\nu E_l}
\delta (E_\nu + E_l -m_\pi). \label{22}\ee where $\theta_c$ is the
Cabibbo angle. Assuming the dispersion relation in Eq. (\ref{4})
(with $m_\nu = 0$) for neutrinos, the integral in Eq. (\ref{22})
evaluates to: \be \frac{E_\nu}{E_\nu + E_l} -\mu \frac{E_\nu +
2E_l}{(E_\nu + E_l)^2} =  \frac{m_\pi^2 - m_l^2}{2m_\pi^2} -\mu
\frac{3m_\pi^2 + m_l^2}{2m_\pi^3}.\label{23}\ee Thus, at the first
order in small quantities, the decay rate takes the form: \be
\Gamma (\pi \rightarrow l \nu)\simeq \Gamma_{\rm B}\left\{1-
2\frac{\mu}{m_\pi} \frac{m_\pi^2 + m_l^2}{m_\pi^2
-m_l^2}\right\}\label{24}\ee where \be \Gamma_{\rm B} =
\frac{G_{\rm F}^2\cos{\theta_c}^2 f_\pi^2m_l^2m_\pi
}{8\pi}\left(1- \frac{m_l^2}{m_\pi^2}\right)\label{25}\ee
indicated the standard Born rate \cite{Okun}. A precise
determination of the $\mu$ parameter again would require the
computation of the radiative corrections (see for example
\cite{radiativep}) but, as in the neutron decay case, we are
interested only in an order of magnitude estimate, so that we will
now proceed as in the previous section. The ratio in Eq.
(\ref{21}) is, then, written as follows: \be R_{e\mu}\simeq
R_{e\mu}^0\left\{1- 2\frac{\mu_e}{m_\pi} \frac{m_\pi^2 +
m_e^2}{m_\pi^2 -m_e^2} + 2\frac{\mu_\mu}{m_\pi} \frac{m_\pi^2 +
m_\mu^2}{m_\pi^2 -m_\mu^2} + \ {\rm radiative \,\,corrections}
\right\}\,,\label{26}\ee where \be R_{e\mu}^0  =
\frac{m_e^2}{m_\mu^2}\,\,\frac{m_\pi^2 - m_\mu^2}{m_\pi^2
-m_\mu^2}\ee is the standard value of the ratio in (\ref{26}).
From the experimental values for the particle masses \cite{PDG}
and the ratio $R_{e\mu}$, we deduce the following constraint on
the $\nu_e$ and $\nu_\mu$ Lorentz-violating parameter: \be
-1.0\mu_e + 3.7\mu_\mu\leq 0.2 \ {\rm MeV}\,.\ee The
evaluation of the rate in the case of the modified dispersion
relation in (\ref{13}) proceeds in an analogous way; after some
algebra we obtain: \be \Gamma (\pi \rightarrow l \nu)\simeq
\Gamma_{\rm B} \left\{1- \frac{\alpha m_\pi}{8M}\left(1-
\frac{m_l^2}{m_\pi^2}\right)\left(3+
5\frac{m_l^2}{m_\pi^2}\right)\right\}\label{29}\,;\ee then the
relation in Eq. (\ref{21}) takes the form: \bea R_{e\mu} &\simeq&
R_{e\mu}^0\left\{1- \frac{\alpha_e m_\pi}{8M} \left(1-
\frac{m_e^2}{m_\pi^2}\right)\left(3 + 5
\frac{m_e^2}{m_\pi^2}\right) \right. \nonumber \\
& & \left. + \frac{\alpha_\mu m_\pi}{8M} \left(1-
\frac{m_\mu^2}{m_\pi^2}\right)\left(3 + 5
\frac{m_\mu^2}{m_\pi^2}\right) + {\rm radiative \,\,corrections}
\right\}\,.\label{30}\eea By using the experimental date \cite{PDG}
we have: \be -3.0\frac{\alpha_e}{M} + 2.5\frac{\alpha_\mu}{M}\leq
186 \ {\rm MeV}^{-1}\,.\ee

\section{Conclusions}

\noindent A possible violation of Lorentz symmetry can manifest into a
variety of process, briefly outlined in the introduction. However,
the energy scales relevant for these processes are usually
extremely large, and the corresponding phenomenology is,
consequently, not accessible by laboratory experiments, so that
only very poor limits on Lorentz violating parameters can be
obtained. A slightly better situation arises in neutrino
phenomenology where, on one hand, some parameters are completely
unconstrained and, on the other hand, very precise experiments are
available. In this paper we have determined the actually most
stringent limits on several Lorentz violating parameters in the neutrino
sector coming from laboratory experiments. In particular we have
focused on the modifications induced by Lorentz violation on the
energy-momentum dispersion relations for $\nu_e$ and $\nu_\mu$ in
two typical cases considered in many different models appeared in
the literature. We have thus reviewed the existing experiments and
found that the most stringent bound on $\nu_e$ parameters comes
from the neutron $\beta$ decay process, while constraints on
$\nu_\mu$ parameters have been obtained by analyzing pion leptonic
decays. The results obtained have been shown in Sections 2 and 3. \\
Those two particular cases of the LV dispersion relation (linear and 
cubic corrections) which we have here studied constitute a starting 
point for a more general phenomenological analysis. Actually, a 
complete theory accounting for Lorentz violation in the neutrino sector, 
is the Standard Model Extension developed in \cite{Mewes}. 
The most general form of the energy-momentum dispersion law for neutrinos, 
including all the violating constants, is obtained in this framework
(see, for example, Eq. (2) in the first paper of Ref.\,\cite{Mewes}). 
A comprehensive analysis of laboratory bounds on these parameters will be 
perfomed in a forthcoming paper. However, although the expected parameter region 
should be, as deduced in the present work, only poorly constrained, 
nevertheless the bounds considered here for the first time indicate
the sensitivity of terrestrial experiments to this kind of
physical problem, and set the starting point for future
experimental investigations on Lorentz symmetry violation.

\

\

\noindent {\bf Acknowledgements}

\noindent The authors are indebted with Prof. E. Recami for very
interesting discussions.

\end{document}